\documentclass[12pt]{iopart}
\usepackage{color}  

\usepackage{soul}

\begin{document}

\title[Fermi equation of state with finite temperature corrections in quantum space-times ...]{Fermi equation of state with finite temperature corrections in quantum space-times approach: Snyder model {vs GUP} case}

\author{Anna Pacho\l$^1$\footnote{Corresponding author}, Aneta Wojnar$^2$}

\address{$^1$ Department of Microsystems, University of South-Eastern Norway, Campus Vestfold, Raveien 215, 3184 Borre, Norway}
\ead{anna.pachol@usn.no}
\vspace{10pt}
\address{$^2$Departamento de F\'isica Te\'orica \& IPARCOS,
	Universidad Complutense de Madrid, E-28040 Madrid, Spain}
\ead{awojnar@ucm.es} 
\vspace{10pt}
%\begin{indented}
%\item[]August 2017
%\end{indented}

\begin{abstract}
We investigate the impact of the deformed phase space associated with the quantum Snyder space on microphysical systems. The general Fermi-Dirac equation of state and specific corrections to it are derived. We put emphasis on non-relativistic degenerate Fermi gas as well as on the temperature-finite corrections to it. Considering the most general one-parameter family of deformed phase spaces associated with the Snyder model allows us to study whether the modifications arising in physical effects depend on the choice of realization. It turns out that we can distinguish three different cases with radically different physical consequences.
\end{abstract}
%
% Uncomment for keywords
\vspace{2pc}
\noindent{\it Keywords}: non-commutative geometry, Snyder model, generalized uncertainty principle, equation of state, stars
%
% Uncomment for Submitted to journal title message
%\submitto{\JPA}
%
% Uncomment if a separate title page is required
%\maketitle
% 
% For two-column output uncomment the next line and choose [10pt] rather than [12pt] in the \documentclass declaration
%\ioptwocol
%
\section{Introduction}
Quantum theory of gravity and the search for phenomenological signatures thereof have been rapidly developing due to the recent rise in experimental precision. The introduction of quantum structure of space-time and the deformation of the associated quantum phase spaces leading to the generalization of the Heisenberg uncertainty principle is growing in importance due to the possible measurable effects emerging from this approach. Generalisations of (the Heisenberg) uncertainty principle (GUPs) have been quite fruitful in providing predictions for quantum gravity effects, without specifying the mathematical structure of the models, although most, see e.g. \cite{GUP1}-\cite{GUP2}, sharing the minimum length scale property, expected to be of the order of the Planck length $L_P\sim \sqrt{\frac{\hbar G}{c^3}}$, see also \cite{GUP3}, \cite{GUP4}, \cite{GUP5}. Nevertheless, the uncertainty relationship between two physical quantities is closely related to their commutation relation, making the deformation of quantum phase space and non-commutative quantum space-times a natural background for GUP theories.
One of the quantum space-time models, introducing non-commutative (NC) coordinates, which can be associated with the generalized uncertainty principle (GUP) framework is the Snyder model \cite{Snyder}. It was proposed in 1947 as the first example of Lorentz-covariant NC space-time, admitting a fundamental length scale. The non-commutativity of the Snyder space-time is encoded in the commutation relation between the space-time coordinates and is proportional to the Lorentz generators of rotations and boosts.  The Lorentz symmetry underlying this space-time is undeformed
at the algebraic level.

In this paper we are interested in the phase space corresponding to the Snyder model, i.e. appropriate deformation of the Heisenberg algebra expressed by the modification of the quantum-mechanical commutation relations between the Snyder  coordinates and momenta. Such deformed algebra is closely related with the modification of the quantum-mechanical uncertainty principle, and thus the physical consequences of the model follow from that algebraic description. Basing on the 'general realization' of the Snyder model, proposed in \cite{BatMelj2}, we consider one-parameter family of deformed phase spaces \cite{AP_symmetry} associated with that model of quantum space-time.

The novelty in our approach, in contrary to GUP literature, is to start with the specific model of the non-commutative space-time, associated with the more general form of the deformed phase space, parametrized by an additional parameter and analyse 
%quantum gravity 
the phenomenological predictions following this deformation.
%of the uncertainty principle tied to this model. 
Considering the one-parameter family of deformed phase spaces, which is linked to the 'general realization' of the Snyder model will also allow us to see if any of the measurable effects pick out one realization (i.e. specific value of the parameter) over the others.

The paper is organized as follows. In the next section we summarize the important details related to the Snyder model, its generalized phase space as well as its relation to quantum symmetry group.
Then we investigate the consequences of the Snyder deformed phase space on microscopic properties of matter, mainly focusing on an equation of state by considering the partition function. Firstly, we derive the general Fermi-Dirac equation of state, with particular emphasis on the non-relativistic degenerate Fermi gas as well as on the temperature-finite corrections to it.
Section 4 is devoted to discussion on the choices of possible realizations and their physical interpretation deduced from the form of the modified equation of state. We finish with short conclusions.

\section{Snyder model and its realizations}
Snyder non-commutative quantum space-time \cite{Snyder} is defined by the following commutation relations: 
\begin{equation}
\lbrack \hat{x}_{\mu},\hat{x}_{\nu}]=i\hbar\beta M_{\mu\nu}  \label{Snyder_orig}
\end{equation}
between the position operators 
$\hat{x}_{\mu}$, and $M_{\mu\nu}$ are the generators of the Lorentz algebra which is the symmetry of this NC space-time. Parameter $\beta $ is the deformation parameter of dimension $\left[\frac{L^2}{\hbar^2}\right]$ that sets the scale of non-commutativity (as $L$ - length is usually associated with the Planck length $L_{p}$). We shall use the notation $\mu,\nu=0,1,\dots ,D$ and $i,j=1,2,\ldots, D$ and the signature is given by $\eta_{\mu\nu}=diag(-,+,\ldots,+)$. The Lorentz symmetry underlying this space-time is undeformed at the algebraic level and is described by the usual Lie algebra relations:
\begin{equation}
\lbrack M_{\mu\nu},M_{\rho\sigma}]=i\hbar(\eta_{\mu\rho}M_{\nu\sigma}-\eta_{\mu\sigma}M_{\nu\rho}+\eta
_{\nu\sigma}M_{\mu\rho}-\eta_{\nu\rho}M_{\mu\sigma}).  \label{MM}
\end{equation}
We also have the cross-commutation relations between Lorentz generators and
Snyder coordinates: 
\begin{equation}
\lbrack M_{\mu\nu},\hat{x}_{\rho}]=i\hbar(\eta _{\mu\rho}\hat{x}_{\nu}-\eta _{\nu\rho}\hat{x}_{\mu}).
\label{Mx}
\end{equation}
Moreover, one can extend the symmetry from Lorentz to the Poincar\'e by considering the usual translation generators $p_\mu$, which commute with each other $[p_\mu,p_\nu]=0$ and satisfy the standard cross-commutation relations with the Lorentz generators:
\begin{equation}
\lbrack M_{\mu\nu},{p}_{\rho}]=i\hbar(\eta_{\mu\rho}{p}_{\nu}-\eta_{\nu\rho}{p}_{\mu}).
\label{Mp}
\end{equation}
The Casimir operator is then defined by $p^2 =\eta^{\mu\nu}p_\mu p_\nu$. However, all the deformation of the Poincar\'e symmetry underlying the Snyder space-time is contained in a (highly) non-trivial co-algebraic
sector \footnote{In the formalism of quantum groups of symmetry associated with the non-commutative space-times, we consider the co-algebraic sector defined by coproducts, antipodes and counits. For the Snyder model, although the quantum group of symmetry is not a Hopf algebra, one can define the co-algebraic sector as follows. All the maps on the Lorentz generators are undeformed (primitive on generators), hence the Lorentz symmetry is truly the undeformed one. But the coproduct on the momenta generators turns out to be deformed and non-coassociative. One can consider different realizations for the non-commuting coordinates leading to the different forms of coproducts on momenta. In this paper, following \cite{BatMelj2}, \cite{AP_symmetry}, the (non-coassociative) coproduct of momenta, up to the linear order in the parameter $\beta $, and the remaining coalgebra maps are defined as follows: 
\begin{eqnarray}
\Delta p_{i} &=&1\otimes p_{i}+p_{i}\otimes 1+    \nonumber\\
&+&\beta \left( \left( \xi-\frac{1}{2}\right) p_{i}\otimes p^{2}+\left( 2\xi-\frac{1}{2}\right) p_{i}p_{k}\otimes p^{k}+\xi\left( p^{2}\otimes
p_{i}+2p_{k}\otimes p^{k}p_{i}\right) \right) +O(\beta ^{2}), \nonumber\\
\epsilon (p_{i}) &=&0,\quad S(p_{i})=-p_{i}.\nonumber
\end{eqnarray}}.

In \cite{BatMelj1} it has been shown that, by using the concept of realizations \footnote{For the precise comparison between the (Heisenberg) realization with the Heisenberg representation and the Hilbert space representation, see e.g. \cite{ABAP_SIGMA}},
there exists infinitely many deformed Heisenberg algebras which correspond to Snyder geometry. By choosing the most general Lorentz-covariant realization for the non-commuting Snyder coordinates  \cite{BatMelj2} (parametrized by $\xi$), the deformation of the quantum-mechanical phase space, up to the linear order in the non-commutativity parameter $\beta$, has the following form \cite{AP_symmetry}:
\begin{eqnarray}
\left[ p_{\mu},\hat{x}_{\nu}\right] &=&-i\hbar\eta _{\mu\nu}\left( 1+\beta \left( \xi-\frac{1}{2}\right) p^{\rho}p_\rho\right)
-2i\hbar\xi\beta p_{\nu}p_{\mu}+O(\beta ^{2}). \label{gen_real_p-x}
\end{eqnarray}
 The original Snyder realization \cite{Snyder}
is recovered for $\xi = 1/2$, giving the usually considered phase space relations:
\begin{equation}
\left[ p_{\mu},\hat{x}_{\nu}\right] =-i\hbar(\eta _{\mu\nu}+\beta p_{\nu}p_{\mu}).\label{c12}
\end{equation}
For $\xi=0$, one obtains the type of realization which can be linked to \cite{Maggiore_GUP1}, \cite{Maggiore_GUP2} \footnote{In \cite{Maggiore_GUP1}, \cite{Maggiore_GUP2} the author arrived at the deformed Heisenberg algebra by investigating the relationship between the generalized uncertainty principle and the quantum deformation of the Poincar\'e algebra without a priori assuming the Snyder model. The commutation relations obtained therein in the finite form can be expanded and compared to the above one given in the first order of $\beta$.} and giving:
\begin{equation}
\left[ p_{\mu},\hat{x}_{\nu}\right] =-i\hbar\eta _{\mu\nu}\left( 1-\frac{\beta }{2}p^{2}\right)+O(\beta ^{2}). \label{c0}
\end{equation}
The so-called
Weyl realization \cite{Weyl_real}, \cite{2003Melj} is obtained for $\xi = 1/6$, with 
\begin{equation}
\left[ p_{\mu},\hat{x}_{\nu}\right] =-i\hbar\eta _{\mu\nu}\left( 1-\frac{\beta }{3}p^{2}\right) -\frac{i\hbar}{3}\beta p_{\mu}p_{\nu}+O(\beta ^{2}).  \label{ph_spWeyl}
\end{equation} 
The first two types (\ref{c12}), (\ref{c0}) have been widely investigated in the context of GUP theories, see e.g. \cite{GUP1,Maggiore_GUP1,Maggiore_GUP2,GUP2,GUP3,ali,ali2,bern,tuna,math,wang}. In this paper we are interested in using the most general one-parameter family of deformed phase spaces (\ref{gen_real_p-x}) corresponding to the Snyder model and investigate the possible measurable effects associated with this deformation. It is worth to point out that the form of (\ref{gen_real_p-x}) agrees with the most general quadratic relativistic GUP, allowing the existence of Lorentz invariant minimum measurable length, considered in \cite{RGUP}. %QuesneTkachuk.
However, since in this paper we are interested in comparing our approach to the one considered in GUP literature, we shall focus only on the spacial part of our deformed phase space (\ref{gen_real_p-x}) and the non-relativistic quantum mechanical picture.
The Heisenberg algebra generated by $\hat{x}_i$ and $p_i$ obeying the commutation relation:
    \begin{eqnarray}
\left[ p_{i},\hat{x}_{k}\right] &=&-i\hbar\delta_{ik}\left( 1+\beta \left( \xi-\frac{1}{2}\right) p^{j}p_j\right)
-2i\hbar\xi\beta p_{i}p_{k}+O(\beta ^{2}). \label{gen_real_p-x_sp}
\end{eqnarray}
following from (\ref{gen_real_p-x}) and parametrized by $\xi$, can be represented on momentum space wave functions $\phi(p)$ with $\hat{x}_i$ and $p_i$ acting as operators \cite{GUP1}:
\begin{equation}\label{x-op}
\hat{x}_{i}\phi(p)=i\hbar\left( \left( 1+\beta \left( \xi-\frac{1}{2}\right) p^{k}p_k\right) 
\frac{\partial }{\partial p_{i}}+2\xi\beta p_{i}p_{j}\frac{\partial }{\partial
p_{j}}+\gamma p_{i}\right) \phi(p),
\end{equation}
\begin{equation}
p_{i}\phi(p)=p_i\phi(p) 
\end{equation}
on the dense domain of functions decaying faster than any power, where $\gamma$ (of dimension $\left[\frac{L^2}{\hbar^2}\right]$) is an arbitrary constant, which does not enter the commutation
relations (\ref{gen_real_p-x_sp}), but affects the definition of the scalar product in momentum space. 
One can check explicitly that this representation obeys the commutation relation (\ref{gen_real_p-x_sp}). 
Further, in order to define symmetric operators:
\begin{equation}
\langle \hat{x}_i\psi,\phi\rangle=\langle\psi,\hat{x}_i\phi\rangle,\qquad \langle p_i\psi,\phi\rangle=\langle\psi,p_i\phi\rangle
\end{equation} 
%(\textcolor{red}{In our notation $\beta \rightarrow \beta \left( c-\frac{1}{2}\right) $ and $\beta ^{\prime }\rightarrow 2c\beta $}) we obtain
%\begin{equation}
%\hat{x}_{i}=i\left( \left( 1+\beta \left( c-\frac{1}{2}\right) p^{2}\right) 
%\frac{\partial }{\partial p_{i}}+2c\beta p_{i}p_{j}\frac{\partial }{\partial
%p_{j}}+\gamma p_{i}\right) 
%\end{equation}
the new inner product in momentum space must be defined as follows \cite{GUP1}, \cite{chang2}:
\begin{equation}
 \langle\psi,\phi\rangle = 
\int \frac{d^D{p}}{(1+\beta(3\xi-\frac{1}{2})p^2)^{\alpha}}\psi^*({p}) \phi({p})
\end{equation}
where $\alpha=\frac{\beta(2\xi+ D\xi-\frac{1}{2})-\gamma}{\beta(3\xi-\frac{1}{2})}$ and is dimensionless.
%$([\beta]=[\gamma]=[\frac{length^2}{\hbar^2}]=\frac{m^2}{m^4 kg^2/s^2}=\frac{s^2}{m^2 kg^2})$.
%$\alpha' = \frac{\gamma -\beta(c-\frac{1}{2})\left(\frac{D-1}{2}\right))}%%%%\beta(c-\frac{1}{2})+2c\beta}$. \\
We also introduce a shortcut notation as $\omega=\beta(3\xi-\frac{1}{2})$
%Let $1-\alpha'=\alpha$ and $\beta(3\xi-\frac{1}{2}))=\omega$. 
then the inner product can be re-written in the following form:
\begin{equation}\label{measure}
    \langle\psi,\phi\rangle = 
     \int \frac{d^D{p}}{(1+\omega p^2)^{\alpha}}\psi^*({p}) \phi({p}).
\end{equation}
%\textcolor{blue}{units: alpha' is 1, omega = $(kg\, m/s)^{-2}$}
The values of $\alpha$ and $\omega$ are closely related with the choice of the realization parameter $\xi$ appearing in the Snyder phase space we consider (cf. (\ref{gen_real_p-x})). For example, for the three distinguished values of parameter $\xi$ discussed above we obtain the following cases:
\begin{itemize}
    \item for the (original) Snyder realization $\xi=\frac{1}{2}$ then $\omega=\beta$ and $\alpha= \frac{\beta(D+1)-2\gamma}{2\beta}$,
    \item for $\xi=0$ we have $\omega=-\frac{1}{2}\beta$ and $\alpha= \frac{\beta+2\gamma}{\beta}$,
    \item for the Weyl realization, $\xi=\frac{1}{6}$ and for this value we do not have the deformation of the measure ($\omega=\beta(3\xi-\frac{1}{2})=0$) and the inner product on the momentum space stays unchanged $\langle\psi,\phi\rangle = 
     \int d^D{p}\psi^*({p}) \phi({p}).$
\end{itemize}
%\textcolor{red}{??moze to usunac?? Due to that fact, we are dealing with a modification to  phase space \cite{chang, paper mentioned in sec3} - the size of each unit cell, being occupied by a quantum state turns out to be also momentum-dependent.}

\section{Equations of state}
In what follows, we would like to derive Fermi equation of state (EoS) resulting from phase space deformations associated with the Snyder NC space. The Fermi gas model is highly significant in the field of physics of stars and substellar objects. Its various forms, such as for instance the well-known polytrope \cite{hore}, polytrope with finite temperature \cite{auddy} are commonly utilized to depict specific regions within the interiors of neutron stars, white dwarf stars \cite{glen}, and non-relativistic stars including pre- and Main Sequence stars \cite{hansen}. It is also popular to model interiors of substellar objects, that is, brown dwarfs and planets \cite{burrows,seager}. %- \textcolor{red}{"pre- and Main" z małej i z dużej litery? pre- odnosi się do sequence, czy do stars? zalezy, czy amerykaie sprawdzaja czy europejczycy. Main Sequence stars uzywam, i w zaleznosci, kto sprawdza, to potem poprawiam}

Let us consider a system of $N$ particles with the energy states $E_i$. %, that is, there are $n_i$ particles in the state $E_i$ \footnote{\textcolor{red}{so $\sum_i n_i = N$? this number does not appear in (23), so does this number matter? / Note that the index $i$ now enumerates different particles and not the spacial part of the coordinates.} \aw{cyba tak, ale nie jestem pewna. wyrzuce $n_i$}}. 
The partition function in the {grand-canonical ensemble} is given as (see e.g. \cite{pad})
\begin{equation}\label{part}
    \mathrm{ln}Z = \sum_i \mathrm{ln}\left[1+az e^{-E_i/k_BT}\right]
\end{equation}
where $T$ is the temperature, $k_B$ Boltzmann constant, $z=e^{\mu/k_BT}$ while $\mu$ is the chemical potential and $a=1$ ($a=-1$) if the particles are fermions (bosons). 

In investigating the effects of the quantum space (\ref{Snyder_orig}) and its phase space deformation (\ref{gen_real_p-x_sp}) on the Fermi equation of state we need to take into account the appropriate modifications in the partition function calculation which are compatible with such deformation. We first note that the deformation of phase space leads to the modified phase space volume\footnote{{To find the Liouville measure one computes the determinant of the symplectic form, and the commutator between coordinates does not contribute to this determinant.}}%, according to the measure deformation (\ref{measure})
, i.e. $(1+\omega p^2)^{-\alpha} d^3xd^3p$ where $\alpha=\frac{\beta(5\xi-\frac{1}{2})-\gamma}{\beta(3\xi-\frac{1}{2})}$
in $D=3$ dimensions.
%partition function etc in Equations of State and Mass-Radius Relations of Quadratic Generalized Uncertainty Principle-modied White Dwarfs with Arbitrary Temperatures
%There are two ways to compute deformed $Z$: working with a standard Heisenberg algebra but with a deformed Hamiltonian or considering the deformed Heisenberg algebra with the non-deformed Hamiltonian. In what follows, we will focus on the second case. 
If we consider a large volume, the summation in the above partition function (\ref{part}) should be replaced by\footnote{{We note that, as shown in} \cite{Chang}, {we can assume that the volume of the phase space evolves in such a way that the number of states inside does not change in time. This holds in the case when the coordinates are non-commutative and the proof of invariance of the weighted phase space volume, indeed relies on the non-commutativity of the coordinates as well.}}
\begin{equation}
      \sum_i\rightarrow \frac{1}{(2\pi \hbar)^3} \int \frac{d^3xd^3p}{(1+\omega p^2)^{\alpha}} 
\end{equation}
Therefore, the partition function in 3 dimensions is given as 
\begin{equation}\label{partition}
     \mathrm{ln}Z =  \frac{V}{(2\pi \hbar)^3}\frac{g}{a}\int \mathrm{ln}\left[1+az e^{-E/k_BT}\right] \frac{d^3p}{(1+\omega p^2)^{\alpha}} \ ,
\end{equation}
where $g$ is a spin of a particle, $V:=\int d^3x$ is the volume of the cell (of the configuration space), while $E=(p^2c^2+m^2c^4)^{1/2}.$ 
As mentioned earlier, the squared physical momentum $p^\rho p_\rho$ is a Casimir invariant of the quantum symmetry of the Snyder model, and the above dispersion relation, related to the Casimir invariant of the Poincar\'e group, stays undeformed. Although, if one considers an effective mass the dispersion relation could be modified \cite{BatMelj2}; we postpone this case for another investigation. 
The thermodynamic variables such as pressure, number of particles, and internal energy are given as, respectively:
\begin{equation}\label{therm}
    P=k_B T\frac{\partial}{\partial V} \mathrm{ln}Z,\;\;\;n=k_B T\frac{\partial}{\partial \mu} \mathrm{ln}Z\mid_{T,V},\;\;\;U=k_B T^2\frac{\partial}{\partial T} \mathrm{ln}Z\mid_{z,V} \ .
\end{equation}
Considering the spherical symmetric case, we can write (\ref{partition}) as 
\begin{equation}\label{partition1}
     \mathrm{ln}Z =  \frac{V}{(2\pi \hbar)^3}\frac{g}{a}\int \mathrm{ln}\left[1+az e^{-E/k_BT}\right] \frac{4\pi p^2 dp}{(1+\omega p^2)^{\alpha}} \ .
\end{equation}

%\subsection{Ideal gas}
%\textcolor{red}{if differs wrt to GR, derive, if not, a short note about that} \aw{maybe it will not be necessary}

\subsection{General Fermi-Dirac equation of state}
In this generalized case, using (\ref{partition1}) and (\ref{therm}) with $a=1$ for fermions and $g=2$ for electrons, we obtain the microphysical description of the system with the Fermi-Dirac distribution $f(E)$, that is,
\begin{equation}
    f(E)=\left(1+z e^{-E/k_BT}\right)^{-1},
\end{equation}
such that the pressure is given by 
\begin{equation}\label{pressuregen}
      P=  \frac{1}{\pi^2 \hbar^3}\int \frac{1}{3}p^3\, _{2}F_{1}\left(\frac{3}{2},\alpha,\frac{5}{2},-p^2\omega\right) f(E) \frac{c^2 p}{E}dp,
\end{equation}
where $_{2}F_{1}$ is the hypergeometric function, while the particle number density and internal energy are
\begin{equation}
    n=  \frac{1}{\pi^2 \hbar^3}\int f(E) \frac{ p^2 dp}{(1+\omega p^2)^{\alpha}} 
\end{equation}
\begin{equation}
    U=  \frac{1}{\pi^2 \hbar^3}\int Ef(E) \frac{ p^2 dp}{(1+\omega p^2)^{\alpha}}.
\end{equation}
For the case when $|\omega p^2| << 1$
\footnote{Note that this condition is satisfied thanks to the fact that $\beta$ is small and it is not a restriction. Nevertheless, we can consider two cases when $\omega>0$ leading to $\xi>\frac{1}{6}$ and $\omega <0$ leading to $\xi<\frac{1}{6}$.}, we can write the pressure as
\begin{equation}
      P=  \frac{1}{\pi^2 \hbar^3}\int \frac{p^3}{3} \left( 
      \sum_{k=0}^{\infty} \frac{ (\alpha)_k\left(\frac{3}{2}\right)_k(-\omega p^2)^k}{ \left(\frac{5}{2}\right)_k k!}
      \right) f(E) \frac{c^2 p}{E}dp 
\end{equation}
while taking into account only the first two terms of the series ({as we consider the NC deformation only up to linear terms in} $\beta$, cf. \ref{gen_real_p-x_sp}. $\beta$ {taken as a small parameter suggests that higher orders will rather not have any significant impact on the physics of stars}), we have
\begin{equation}
      P=  \frac{1}{\pi^2 \hbar^3}\int \left(\frac{p^3}{3} -\frac{\omega\alpha p^5}{5} \right) f(E) \frac{c^2 p}{E}dp \ .
\end{equation}
We note that in our approach $\alpha\omega$ depends on the realization of the deformation of the phase space $\xi$ in (\ref{gen_real_p-x_sp}), hence we can consider cases when $\alpha\omega=0$, $\alpha\omega>0$ or $\alpha\omega<0$. We discuss the implications of this in detail in Sec. 4.

\subsection{Non-relativistic degenerate Fermi gas}

Before considering a simplified case (but keeping the general realization), let us rewrite the general Fermi-Dirac pressure (\ref{pressuregen}) in terms of energy for the non-relativistic electrons, that is, $E\approx\frac{p^2}{2m_e}$:
\begin{equation}
      P=  \frac{1}{\pi^2 \hbar^3}\int \frac{1}{3}(2m_eE)^\frac{3}{2}\, _{2}F_{1}\left(\frac{3}{2},\alpha,\frac{5}{2},-2Em_e\omega\right) f(E) dE.
\end{equation}
For the case when $2m_e|\omega E| << 1$, we can write the pressure as
\begin{equation}\label{pres2}
      P=  \frac{1}{3\pi^2 \hbar^3}\int  \left( (2m_eE)^\frac{3}{2} -\frac{3\alpha\omega}{5} (2m_eE)^\frac{5}{2}\right) f(E) dE.
\end{equation}
We will focus now on a specific type of matter which has a very special interest in the stellar physics: degenerate gases. They are used to describe dense cores of stellar and substellar objects as well as degenerate matter in compact stars. In the toy model case which we are going to use here, all states with energy being less than Fermi energy level are occupied while all states with higher energy are empty at the absolute temperature, that is, $T\rightarrow0$, and the chemical potential $\mu$ becomes the Fermi energy $E_F$. For such a case, the Fermi-Dirac distribution is a step function
\[ f(E)=  \left\{
\begin{array}{ll}
      1 & \mbox{if } E\leq E_F \\
      0 & \mbox{otherwise.} \\
\end{array} 
\right. \]
%\begin{equation}
%f(E)=
%    \begin{cases}
%        1  \mbox{if } E\leq E_F \\
%        0  \mbox{otherwise.}
%            \end{cases}
%\end{equation}
Due to that fact, the integration in (\ref{pres2}) is taken up to the Fermi energy $E_F$ such that
\begin{equation}\label{pres3}
      P_{T\rightarrow0}=  \frac{2}{5}vE_F^\frac{5}{2} \left(1 -\frac{3\alpha\omega}{7} (2m_e) E_F\right) ,
\end{equation}
where we have defined $v=\frac{(2m_e)^\frac{2}{3}}{3\pi^2 \hbar^3}$. The pressure $P_{T\rightarrow0}$ becomes smaller when $\alpha\omega>0$ while it increases for $\alpha\omega<0$.

Let us use the definition of the measure of electron degeneracy 
\begin{equation}\label{degeneracy}
\psi=\frac{k_{B} T}{E_{F}}=\frac{2 m_{e} k_{B} T}{\left(3 \pi^{2} \hbar^{3}\right)^{2 / 3}}\left[\frac{\mu_{e}}{\rho N_{A}}\right]^{2 / 3}\equiv u^{-1}  k_{B} T \left[\frac{\mu_{e}}{\rho }\right]^{2 / 3}
\end{equation}
in the equation (\ref{pres3}) to write
\begin{equation}\label{pres4}
    P_{T\rightarrow0} =\frac{2}{5}vu^\frac{5}{2}\left(\frac{\rho}{\mu_e}\right)^\frac{5}{3}\left[
    1-\frac{3u}{7} \alpha\omega (2m_e)\left(\frac{\rho}{\mu_e}\right)^\frac{2}{3}
    \right]
\end{equation}
where $u=(3 \pi^{2} \hbar^{3} N_A)^\frac{2}{3}/2m_e$. It can
be rewritten as a mixture of two polytropes %with $n=3/2$ and $n=3/4$
\begin{equation}%\label{}
    P_{T\rightarrow0}= K_1 \rho^{\Gamma_1} - \alpha\omega K_2 \rho^{\Gamma_2}
\end{equation}
where $K_1= \frac{2}{5}vu^\frac{5}{2}\mu_e^{-\frac{5}{3}}$, $\Gamma_1 = 5/3$ and $K_2= \frac{12}{35}vu^\frac{7}{2}m_e\mu_e^{-\frac{7}{3}}$, $\Gamma_2 = 7/3$. Notice that a similar modification of the polytrope EoS also happens in the case of modified gravity \cite{kim}, where the additional term appears because of the gravitational backreaction on the particles.

Let us notice that using again the definition (\ref{degeneracy}) to the last term we can rewrite this EoS as
\begin{equation}\label{pol1}
     P_{T\rightarrow0}= K_1 \rho^{\Gamma_1} \left(1-\frac{6}{7}m_e \alpha\omega\frac{k_BT}{\psi}\right) = \tilde K  \rho^{\Gamma_1},
\end{equation}
where $\tilde K := K_1  \left(1-\frac{6}{7}m_e \alpha\omega\frac{k_BT}{\psi}\right)$. We see that non-commutativity introduces the electron degeneracy term %\footnote{\textcolor{red}{Does it matter that $\alpha\omega>0$, so that we get minus between two terms? If yes, then we can make some comments about the distinguished realization $\xi=0.1$ for which the sign changes. E.g. is there interpretation for the term $(1+\frac{6}{7}...)$?}} 
which usually appears when we consider corrections from the finite temperature, see e.g. \cite{auddy} and discussion in the next section \ref{tempdep}. Note that we are also dealing with an implicit dependence on the temperature $T$. If $\psi=1$, we are dealing with a non-degenerate matter, while $\psi\rightarrow0$ with the degenerate one, increasing by this the significance of the quantum deformation term. In general, $\psi$ is a function of time. Therefore, we can expect that objects with degenerate matter, such as compact or contracting astrophysical bodies, can be in principle used to test theories with modifications given by the non-commutativity, in a similar manner as it was performed in modified gravity or dark matter models \cite{w1,w2,leane,benito,w3,w4,Guerrero,Kozak,Gomes}.

\subsection{Temperature-finite corrections to the non-relativistic Fermi gas}\label{tempdep}
We can also consider a more realistic situation, that is, when the temperature is finite. Playing a bit\footnote{The procedure we have used here is nicely presented in the appendix A of \cite{auddy}.} with the general form of pressure (\ref{pressuregen}), one may rewrite it as
\begin{eqnarray}
    P&=& \frac{2}{5} v\mu^\frac{5}{2} \left(1 -\frac{3\alpha\omega}{7} (2m_e) \mu\right) -\frac{1}{8} \frac{v}{k_B T} \mu^\frac{3}{2}\mathrm{ln}[1+e^{-\mu/{k_BT}}] \nonumber \\
   &+&\frac{3}{4} \frac{v}{(k_B T)^2} \mu^\frac{1}{2}\mathrm{Li}_2[e^{-\mu/{k_BT}}]
   \left(1 - \frac{\alpha\omega}{10} (2m_e) \mu\right) \nonumber \\
   &-& \frac{3}{4} \frac{v}{(k_B T)^3} \mu^{-\frac{1}{2}}\mathrm{Li}_3[e^{-\mu/{k_BT}}]
    \left(1 - \frac{12\alpha\omega}{10} (2m_e) \mu\right) + ...\,,
\end{eqnarray}
where $Li_s$ are the polylogarithm functions of different orders $s$. As the gas becomes increasingly degenerate, the polylogs exhibit an exponential decrease, therefore further terms turn out to be insignificant in the above expression.

Analogously, as in the previous case, we can rewrite it in terms of the electron degeneracy. Keeping the EoS up to the second order in $\psi$ and considering again $\mu\approx E_F$ one gets
\begin{eqnarray}\label{presT}
    P&=& \frac{2}{5}vu^\frac{5}{2}\left(\frac{\rho}{\mu_e}\right)^\frac{5}{3}\left[
    1  -\frac{5}{16} \psi \mathrm{ln}[1+e^{-1/\psi}] 
    + \frac{15}{8}\psi^2\mathrm{Li}_2[e^{-1/\psi}]
    \right.  \nonumber \\
    &-& \left. 3m_e \alpha\omega\left( \frac{2}{7}\frac{k_BT}{\psi}
    +\frac{1}{8}\frac{\psi}{(k_BT)^3}\mathrm{Li}_2[e^{-1/\psi}] \right)
    \right],
\end{eqnarray}
which again can be written as the polytropic equation of state with with the polytropic index $\Gamma=5/3$ and 
\begin{equation}
  p=   K  \rho^{\Gamma}
\end{equation}
where the polytropic "parameter" is
\begin{eqnarray*}
    K = \frac{2}{5}vu^\frac{5}{2}\mu_e^{-\frac{5}{3}}\left[
    1  -\frac{5}{16} \psi \mathrm{ln}[1+e^{-1/\psi}] 
    + \frac{15}{8}\psi^2\mathrm{Li}_2[e^{-1/\psi}]
    - 3m_e \alpha\omega C(\psi)
    \right]
\end{eqnarray*}
with $C(\psi,T)=\left( \frac{2}{7}\frac{k_BT}{\psi}
    +\frac{1}{8}\frac{\psi}{(k_BT)^3}\mathrm{Li}_2[e^{-1/\psi}] \right)$.

\section{Discussion}

Let us firstly discuss the physical interpretation of the terms introduced to the Fermi EoS by the non-commutativity. Note that the parameter $K$ appearing in the polytropic forms of EoS derived in the previous section (equation (\ref{pres3}) or (\ref{pres4})) is related to the bulk modulus or incompressibility
\begin{equation}
    B= \frac{dP}{d \mathrm{ln}\rho},
\end{equation}
describing properties of an isotropic material, for example crystallized cores of white dwarf stars or terrestrial planets. It can be also written with respect to the shear modulus and elastic constants, appearing in the Hooke's law \cite{poirier}. For our polytropic form, the bulk modulus is then
\begin{equation}\label{bulk}
    B=K\Gamma\rho^\Gamma,
\end{equation}
which is modified due to the modifications in $K$ provided by the deformation parameters. For incompressible solids $B=\infty$ while for infinitely compressible one $B=0$. For example, for our simplified case (\ref{pres3})
$$K=1-\frac{6}{7} \alpha\omega m_e  E_F$$
we will deal with infinitely compressible solid ($B=0$) when $\alpha\omega>0$ while $\alpha\omega<0$ provides the incompressible counterpart. We shall now consider those cases in more detail.
$$***$$
In the equations obtained in the previous sections, we can notice that the combination $\alpha\omega$ (more precisely the sign of this term) influences the physical interpretation of the modifications introduced by the non-commutativity.
The deformation parameter $\beta$ and the parameter $\xi$ generalising the deformation of the phase space (\ref{gen_real_p-x}) are both included in this combination. For any dimension $D$ we have $
\alpha\omega = \beta(2\xi+D\xi-\frac{1}{2})-\gamma.
$
Since we are interested in $D=3$, this formula reduces to 
\begin{equation}\label{par}
\alpha\omega = \beta(5\xi-\frac{1}{2})-\gamma.
\end{equation}
Let us recall that $\gamma$ is an arbitrary constant, which does not enter the commutation relations, but affects the definition of the scalar product in momentum space. When $\beta\to 0$ classical (undeformed) case, i.e. no presence of non-commutative geometry, we have $\alpha\omega=-\gamma$. Considering $\gamma\neq 0$ would lead to non-trivial representation of commutative coordinates in (\ref{x-op}) and such (unusual) choice of representation could possibly lead to some modifications of physical properties. Therefore, to recover the correct undeformed (commutative) space-time limit we should require $\gamma=0$. So let us assume that $\gamma=0$ from now on.

In Sec. 3.1, we have assumed that $|\omega p^2|<<1$ to allow for the expansion of the hypergeometric function in (\ref{pressuregen}). This is not a restriction as the order of magnitude of $\beta$ in $\omega=\beta(3\xi-\frac{1}{2})$ ensures this condition and we are free to consider the cases with any sign of $\omega$. Therefore, we can discuss various sign options for the $\alpha\omega$ term and have explicit dependence of the physical properties on the choice of realization ($\xi$) in (\ref{gen_real_p-x_sp}). That is (considering even the simplest case with $\gamma=0$) we have: $\alpha\omega=\beta(5\xi-\frac{1}{2}) =0 $ when $\xi=\frac{1}{10}$, 
$\alpha\omega >0 $ when $\xi> \frac{1}{10}$
%then the modification in phase space induces a reduction of the available number of microstates, which holds the thermodynamic properties and hence decreases the internal energy and pressure of the system. 
and   
$\alpha\omega < 0 $ when $\xi < \frac{1}{10}$. When the term $\alpha\omega$ changes sign the physical interpretation changes radically. We are now in position to discuss all these cases in detail.

\subsection{$\alpha\omega=0$}
Firstly, we should notice that we can distinguish the value of the parameter $\xi=\frac{1}{10}$ which gives $\alpha\omega=0$. Since the combination $\alpha\omega$ is responsible for introducing modified physical properties, this choice of realization results in no modification to the considered physical effects up to the linear order in the deformation parameter $\beta$.
%the phase space is deformed and is of the form (\ref{gen_real_p-x_xi0.1}. 
%therefore the phase space deformation \textbf{increases the available number of microstates} and increases the internal energy and pressure of the system.
Using this distinguished value of $\xi$ we get the following form of the deformed phase space:
 \begin{eqnarray}
\left[ p_{i},\hat{x}_{k}\right] &=&-i\hbar\delta_{ik}\left( 1- \frac{2}{5}\beta p^{j}p_j\right)
-i\hbar\frac{1}{5}\beta p_{i}p_{k}+O(\beta ^{2}). \label{gen_real_p-x_xi0.1}
\end{eqnarray}
Note that the choice of the realization $\xi=\frac{1}{10}$ will automatically enter in the coproduct of Snyder momenta (see footnote on page 3),  hence will affect the quantum composition law for the momenta.

Let us also point out that for $\xi=\frac{1}{6}$ there is no modification in physical properties of the system, due to the fact that for this value $\omega=0$ and there is no modification in the measure (\ref{measure}).

However, these two cases differ significantly as for $\xi=1/6$ we have no NC corrections in the effects we considered due to the fact that the measure stays undeformed. While for $\xi=1/10$ we have no NC corrections in the effects we considered up to the first order in the deformation parameter. The modifications may appear in the higher orders for this realization.
Although, for both of these numerical values of $\xi$ the space-time and the phase space both stay deformed.

\subsection{$\alpha\omega>0$}
The condition $\alpha\omega>0$ picks out the deformed phase spaces (\ref{gen_real_p-x_sp}) for which $\xi>1/10$. 
This includes $\xi=1/2$ which is one of the most considered cases in the literature. 
%Before putting a numerical bound on parameters because of our previous derivations, let us firstly discuss them. 

Firstly, we notice that in this case one deals with weaker electron degeneracy pressure (\ref{pres3}) which means that one deals with a more compact object with respect to the commutative case. Moreover,
because of the condition $\alpha\omega>0$, we can obtain
its critical value 
\begin{equation}\label{crit}
   (\alpha\omega)_{\mbox{crit}}=\frac{7}{6E_Fm_e},
\end{equation}
for which pressure becomes zero, or, in other words, we deal with infinitely compressible material.
%which we can again rewrite in terms of density $\rho$ using the definition of (non-modified in our current approach) Fermi energy (\ref{degeneracy}. 
We can see that the choice of the realization (value of $\xi$) assures the condition $\alpha\omega>0$ for which the critical value exists.

To put a numerical bound on $(\alpha\omega)_{\mbox{crit}}$, let us take a value of Fermi energy of a typical white dwarf star, $E_F\sim3$ MeV, providing the critical value as $(\alpha\omega)_{\mbox{crit}}\sim2.66\times10^{43}$ (kg\, m/s)$^{-2}$. So for instance, in the realization with $\xi=\frac{1}{2}$ the critical value for the non-commutativity parameter $\beta_{\mbox{crit}}\sim 10^{43}$.
Therefore, from this very simplified analysis, we can put the bound
  \begin{equation}
       \beta_0< 4.27\times 10^{44}
    \end{equation}
where we have used a more common notation, that is, $\beta_0=\beta M^2_Pc^2$ ({dimensionless}).

Our upper bound is consistent with the newest results on the upper bound obtained by comparing the orbital angular momentum acquired after light is lensed by a GUP-modified rotating black hole with experimental data acquired for M87, that is, $\beta_0\sim10^{78}$ \cite{tam}. In \cite{dag2}, the graviton and photon speeds in a GUP-modified curved space-time were compared with
speeds obtained from gravitational wave events GW150914 and GW190521. An upper bound $\beta_0<2.56\times10^{60}$ was provided in the case of taking into account the GUP modifications in the graviton speed only, while more stringent bound $\beta_0<28.83\times10^{35}$ was given when the modification was considered in the photon speed. From computing the Hawking temperature for a Schwarzschild black hole, the authors \cite{sca} obtained $\beta_0\sim10^{10}$. On the other hand, the scanning tunneling microscopy experiments provided a few other bounds, that is, the bound $\beta_0<10^{50}$ was provided by the accuracy of measurements of Landau levels while $\beta<10^{36}$ from the
accuracy in precision measurements of the Lamb shift of hydrogen atom \cite{das}. Moreover, the bound $\beta_0<10^{21}$ is required in order to add a GUP induced
current up to the charge of just one electron. In \cite{brau} the energy spectrum of the gravitational quantum well modified by a first order perturbation of the parameter $\beta$ was compared to the energy spectrum values obtained from the GRANIT experiments, providing $\beta_0<10^{34}$. Considering the fact that number density can only be positive, $\beta_0<10^{13}$ was obtained in \cite{hari}, while analyzing the harmonic oscillator energy levels \cite{chang2} provided $\beta_0<10^{19}$.

\subsection{$\alpha\omega<0$}
%\aw{nowy case, ale sprawdz prace z lower bound}
The condition $\alpha\omega<0$ requires the form of the deformed phase space (\ref{gen_real_p-x_sp}) for which $\xi<1/10$, and includes the case $\xi=0$.

If $\alpha\omega$ is negative, the non-commutative correction makes the solid more incompressible, that is, the astrophysical object is less compact in comparison to the non-modified case. Since the sign in (\ref{pres3}) changes, we do not deal anymore with critical values.

Let us notice that any modification in the equation of state will have a non-trivial effect on a mass and radius of stars. There exist a few well-known limiting masses, such as for example the theoretical Buchdahl's bound \cite{buh} or more empirical methods, allowing to test the theoretical models against observations, such as for instance, minimum Main Sequence mass \cite{burrows,Guerrero,sakstein,olmo} or critical masses/temperatures for Hayashi tracks \cite{w1}. It would be interesting to investigate these ideas in detail in the future.
%We will leave working on those ideas for our future projects.

\section{Conclusions}
The purpose of the present paper was to examine the influence of different parametrizations of the deformed phase space arising from the Snyder model on the microphysical system. Before doing so, we firstly recalled the basic notions and properties of the Snyder model, and related possible choices of the realizations to the existing generalized uncertainty principle frameworks.

Subsequently, considering the most general deformation of the quantum-mechanical phase space in 3D associated with the Snyder model, we derived a general Fermi-Dirac equation of state resulting from the mentioned deformation. The modified terms are governed by $\alpha\omega$, which is a combination of the deformation parameter $\beta$, choice of realization $\xi$ and arbitrary constant $\gamma$ related to the representation of the coordinates in the momentum space. 
As particular examples of the Fermi-Dirac equation of state, we have also derived corrections to the usual polytropic equation of state (that is, for the case when the temperature of the system $T\to0$ and one deals with a non-relativistic degenerate Fermi gas) (\ref{pres4}), as well as we computed the temperature finite corrections to the non-relativistic Fermi gas (\ref{presT}). In both cases, we kept the the most general realization picture.

We observed that the terms introduced by the deformation have a clear physical interpretation. Firstly, as expected, because of modification of the phase space (the size of each unit cell which is occupied by a quantum state) one deals with an increase, $\alpha\omega<0$, or decrease, $\alpha\omega>0$ response from the fermionic fields against attempts at squeezing, for example by the gravitational attraction. Notice that similar term appears in modified gravity \cite{kim} and similar corrections are expected in other models of gravity \cite{w1,wfermi}. Therefore, we should not be surprised to see that those terms (\ref{pol1}) can be written with respect to the electron degeneracy parameter, $\psi\sim T/E_F$, whose value informs how much a given object is degenerate. In the solid state physics language, the bulk modulus (incompressibility) of isotropic materials, such us for example carbon or iron, acquires an extra term, making the material more or less compressible, depending on the sign of $\alpha\omega$. Therefore, we expect to observe interesting effects when studying properties of the white dwarf stars' \cite{kalita}, brown dwarfs' \cite{benito}, or Earth's cores \cite{olek}, respectively. Let us notice, however, that the electron degeneracy parameter depends on time \cite{auddy}, therefore, if one wishes to use any astrophysical bodies to constrain the parameters, one should be more interested in very old astrophysical objects as the effects of non-commutativity will be more pronounced \cite{benito,Kozak}. 

Having this interpretation in mind, since the sign of $\alpha\omega$ is governed by the choice of the realization leading to different deformed phase spaces, we have distinguished particular choices: when $\xi=1/10$ leading to $\alpha\omega=0$, when $\xi>1/10$ giving $\alpha\omega>0$ and $\xi<1/10$ providing $\alpha\omega<0$. Ideally, depending on the measurable physical effects we would be able to determine the specific type of realization detected or preferred.

Moreover, using the simple Fermi gas equation of state (\ref{pres3}) we obtained a critical value for $(\alpha\omega)_{\mbox{crit}}\sim2.66\times10^{43}$ (kg\, m/s)$^{-2}$.

{In the present work, we have focused on the Snyder model, with its similarities of the generalized uncertainty principle allowing us to make direct comparisons with the results available in the GUP literature.
However, one can expect that the common feature of other non-commutative space-times, which lead to the deformation of the phase space, would be the modification of the inner product and consequently the modification of the measure of the phase space volume (see e.g.} \cite{Maggiore_GUP2}, \cite{hari}).{ The modification in the measure, will in turn modify the partition function leading to the corrections arising, for example, in the pressure, number of particles, and internal energy depending on the non-commutativity parameters used in the deformation of the space-time. Going further, one obtains similarities in the form of corrections which may arise from different quantum gravity approaches, such as modified gravity, see e.g.} \cite{APAW2}. {
Therefore similar effects, as considered here, should be relevant for other non-commutative models, and possibly other quantum gravity approaches and would be an interesting topic for further study.}

\subsection*{Acknowledgments}
%The author(s) would like to acknowledge the contribution of the COST Action CA18108.
%For works co-authored by at least two WG/MC members from at least two different countries participating to the Action:

The authors would like to acknowledge networking support by the COST Action CA18108 and STSM Grant No. E-COST-GRANT-CA18108-21665909. AP has been supported by the Polish National Science Center (NCN), project UMO-2022/45/B/ST2/01067. AW acknowledges financial support from MICINN (Spain) {\it Ayuda Juan de la Cierva - incorporac\'ion} 2020.

\end{document}